\documentclass[aps,prd,twocolumn,groupedaddress,floatfix]{revtex4}
\usepackage{epsfig}
\usepackage{dcolumn}
\usepackage{amsmath}
\def\Journal#1#2#3#4{{#1} {\bf#2}, #3 (#4)}

\def\NPA{{\rm Nucl. Phys.} A}
\def\NPB{{\rm Nucl. Phys.} B}
\def\PLB{{\rm Phys. Lett.}  B}
\def\PRL{\rm Phys. Rev. Lett.}
\def\PRD{{\rm Phys. Rev.} D}
\def\PRC{{\rm Phys. Rev.} C}

\def\epp{\epsilon^{\prime}}
\def\ep{\epsilon}

\def\la{\langle}
\def\ra{\rangle}

\def\be{\begin{equation}}
\def\ee{\end{equation}}
\def\bea{\begin{eqnarray}}
\def\eea{\end{eqnarray}}

\begin{document}
\title{Radiative Scalar Meson Decays   
in the Light-Front Quark Model}   
\author{ Martin A. DeWitt$^{a}$, Ho-Meoyng Choi$^{b}$
and Chueng-Ryong Ji$^{a}$\\
$^a$ Department of Physics, North Carolina State University,
Raleigh, NC 27695-8202\\
$^b$ Department of Physics, Kyungpook National University,
     Daegu, 702--701 Korea}
\begin{abstract}
We construct a relativistic $^3P_0$ wavefunction
for scalar mesons within the framework of light-front quark model(LFQM).  
This scalar wavefunction is used to perform
relativistic calculations of absolute
widths for the radiative decay processes $(0^{++})\to\gamma\gamma$,
$(0^{++})\to\phi\gamma$, and $(0^{++})\to\rho\gamma$ which incorporate 
the effects of glueball-$q\overline{q}$ mixing.  The mixed physical states 
are assumed to be $f_0(1370)$, $f_0(1500)$, and $f_0(1710)$ for which the 
flavor--glue content is taken from the mixing calculations of other works.  
Since experimental data for these processes are poor, our results are 
compared with those of a recent non-relativistic model calculation. We 
find that while the
relativistic corrections introduced by the LFQM reduce the magnitudes of 
the decay
widths by 50--70\%, the relative strengths between different decay 
processes are
fairly well preserved. We also calculate decay widths for the processes
$\phi\to(0^{++})\gamma$ and $(0^{++})\to\gamma\gamma$ involving the light 
scalars
$f_0(980)$ and $a_0(980)$ to test the simple $q\bar{q}$ model of these 
mesons. Our results of $q\bar{q}$ model for these processes 
are not quite consistent with well-established data, further 
supporting the idea
that $f_0(980)$ and $a_0(980)$ are not conventional $q\bar{q}$ states.
\end{abstract}
\maketitle

\section{Introduction}
As is well known, the assignment of the scalar($J^{PC}=0^{++}$) 
$q\bar{q}$ states has long been an enigma in hadron 
spectroscopy. Unlike the elegant, ideally mixed vector and
tensor multiplets, it is still controversial which are the 
members of the expected $L=S=1$ $q\bar{q}$ multiplet since 
there are now too many $0^{++}$ mesons observed in the 
region below 2 GeV for them all to be explained naturally  
within a $q\bar{q}$ picture\cite{data}. 
For example, 2 isovector($IJ^{PC}=1 \: 0^{++}$)[$a_{0}(980)$, $a_{0}(1450)$] 
and 5 isoscalar($0 \: 0^{++}$)[$f_{0}(600)$(or $\sigma$),
$f_{0}(980)$, $f_{0}(1370)$, $f_{0}(1500)$, $f_0(1710)$]   
states have been reported by the Particle Data Group\cite{data}. This has
led to the suggestion that not all of them are $q\bar{q}$ states.
The main reason for this situation is that around the relevant mass
region there exist other kinds of particles such as $K\bar{K}$ 
molecules\cite{Isgur,godfrey,barnes}, glueballs\cite{Amsler,Anisovich}, 
four-quark($qq\bar{q}\bar{q}$) systems\cite{Jaffe}, or hybrids.

Interpreting the structure of each of the known scalars has proven to be
a fairly controversial endeavor.  Take, for example, the light scalars
({\it i.e.} those below 1 GeV).  Due to some of the difficulties
associated with $f_0(980)$ and $a_0(980)$---{\it{e.g.}} the
strong couplings to $K\bar{K}$ in spite of their masses being at the $K\bar{K}$
threshold, and the large discrepancies of 
$\pi\pi$\cite{godfrey},$\gamma\gamma$\cite{barnes} 
and $\phi$-radiative decay\cite{kumano} widths between 
non-relativistic (NR) quark model predictions and 
experimental data---Weinstein and Isgur\cite{Isgur} proposed the isospin
$I=0 \: f_{0}(980)$ and the $I=1 \: a_{0}(980)$ within the NR potential 
model as the ``$K\bar{K}$ molecules".  However, a more conventional 
interpretation for these
states has been given by T\"{o}rnqvist and Roos\cite{Nils} who analyzed 
the data on
the $f_{0}(980),f_{0}(1370),a_{0}(980)$ and $a_{0}(1450)$ as unitarized 
remnants of $q\bar{q}$ $1^{3}P_{0}$ states with six parameters and
theoretical constraints including flavor symmetry, the OZI rule, the
equal-spacing rule for the bare $q\bar{q}$ states, unitarity,
and analyticity. In this work, the authors concluded that the 
$f_{0}(980)$ and the $f_{0}(1370)$ are two manifestations of the same 
$s\bar{s}$, while the $a_{0}(980)$ and 
the $a_{0}(1450)$ are two manifestaions of the same $u\bar{d}$ state. 

The interpretation of the structures of some of the heavier scalars
has been somewhat less controversial.
For example, $f_0(1370)$ and $a_0(1450)$ are most
commonly interpreted as $q\bar{q}$ states, even though the flavor assignments
for each are still unclear.  Also, there appears to be general agreement on
identifying $f_0(1500)$ as a glueball, possibly mixed with
$n\bar{n}=(u\bar{u}+d\bar{d})/\sqrt{2}$ and $s\bar{s}$.  This interpretation
has followed both from lattice QCD which, in the quenched approximation, predicts
that the lightest glueball has $J^{PC}=0^{++}$ and a mass of
1.55--1.74 GeV \cite{Bali,Weingarten}, as well as from the fact that
$f_0(1500)$ decays strongly into $\pi\pi$ but not into $K\bar{K}$.

Most recently, Close and T\"{o}rnqvist\cite{CT2002} have proposed a scheme which sorts the
light scalars into two distinct nonets: one nonet above 1 GeV and another below
1 GeV, with different physics operating in each.
The nonet above the 1 GeV threshold is comprised of the $q\bar{q}$ states mixed with
the scalar glueball.  The glueball's presence is inferred from the overpopulation of
isoscalars in this mass region.  The nonet below 1 GeV is made
up of $qq\bar{q}\bar{q}$ and meson-meson molecules.  As such, $f_0(980)$ and
$a_0(980)$ can be thought of as superpositions of four-quark states and $K\bar{K}$
molecules.  The authors of Ref.~\cite{CT2002} demonstrate that such a scheme involving two
scalar nonets can be described using two coupled linear sigma models.

In this paper, we investigate various radiative scalar meson decays
which could provide important clues on the internal structures of these
states\cite{kumano,Nuss}.  For the calculations, we employ the light--front quark
model (LFQM) which has been used very successfully in the past to compute decay rates
for pseudoscalar, vector, and axial--vector mesons \cite{CJ1,CJ2,jaus96}.
Extending the model to include scalar states mainly involves the construction
of a new $^3P_0$ light--front model wavefunction.  In section II, we give the detailed
form of this wavefunction and explain how the model parameters are obtained.
This new scalar wavefunction is used to study the radiative decays
of the heavy scalars $f_0(1370)$, $f_0(1500)$, and $f_0(1710)$, as well
as the light scalars $a_0(980)$ and $f_0(980)$.

In the case of the heavy scalars,
we adopt the scheme of Close and T\"{o}rnqvist in which $f_0(1370)$,
$f_0(1500)$, and $f_0(1710)$ are considered to be mixtures of $n\bar{n}$, $s\bar{s}$,
and $gg$. The flavor--glue content of each state is taken from mixing analyses done
by Lee and Weingarten \cite{LW1999} and by Close and Kirk \cite{CK2001}.  Taken
together, these works provide mixing amplitudes for three distinct cases of the
scalar glueball mass:  (1) a heavy glueball ($M_{n\bar{n}}<M_{s\bar{s}}<M_{gg}$),
(2) a medium weight glueball ($M_{n\bar{n}}\lesssim M_{gg}<M_{s\bar{s}}$), and
(3) a light glueball ($M_{gg}<M_{n\bar{n}}<M_{s\bar{s}}$).  The details of these
three mixing scenarios are outlined in section III.
In the case of the light scalars, $a_0(980)$ and
$f_0(980)$ are assumed to be conventional $q\bar{q}$ states.  The flavor
content of $a_0(980)$ is then $(u\bar{u}-d\bar{d})/\sqrt{2}$, and
that of $f_0(980)$ is some superposition of $n\bar{n}$ and $s\bar{s}$.
Rather than attempting to determine the degree of mixing for $f_0(980)$,
it suffices to examine the two ideally mixed cases: $f_0(980)=n\bar{n}$
and $f_0(980)=s\bar{s}$.

In section IV, the general forms of the $Q^2$--dependent transition form factors
for the processes $(1^{--})\to(0^{++})\gamma^*$, $(0^{++})\to(1^{--})\gamma^*$, and
$(0^{++})\to\gamma\gamma^*$ are derived.  In the limit as $Q^2\to0$, these
form factors yield the decay constants for the real photon processes which can
then be used to compute the corresponding decay widths.  In section V--A, we
present our numerical results for the heavy scalars.  This includes the form factors
and decay widths for the specific processes $f_0\to\phi\gamma$, $f_0\to\rho\gamma$,
and $f_0\to\gamma\gamma$.  Our results for the light scalars involved in the
processes $\phi\to f_0(a_0)\gamma$ and $f_0(a_0)\to\gamma\gamma$ are given in
section V--B.  A summary of the paper's salient points and a brief discussion of
our intended future work is given in section VI. 
In the Appendix, the explicit form of the trace used in section IV.A is 
presented.
\section{The model wavefunctions}
One of the popular quark models in the light-front formalism is the invariant
meson mass(IM) scheme\cite{Jaus,Chung} in which the invariant meson mass 
square $M_{0}^{2}$ is given by
\bea{\label{eq:1}}
M^{2}_{0}&=& \sum^{2}_{i}\frac{{\bf k}^{2}_{i\perp } + m^{2}_{i}}{x_{i}}.
\eea
In our analysis, we will only consider the light-meson
sector($u,d$, and $s$ quarks) with equal quark and anti-quark
masses($m_{q}=m_{\bar{q}}$).

The light-front $q\bar{q}$ bound-state wavefunction of the scalar($^{3}P_{0}$)
and vector($^{3}S_{1}$) mesons can be written in the following covariant form
\bea{\label{eq:2}}
\Psi_{M}(x_{i},{\bf k}_{i\perp},\lambda_{i})&=&
\bar{u}_{\lambda_{q}}(p_{q})\Gamma_{M}v_{\lambda_{\bar q}}(p_{\bar q})
\phi_{M}(x_{i},{\bf k}_{i\perp})\nonumber\\
&=&{\cal R}^{M}_{\lambda_{q}\lambda_{\bar q}}(x_{i},{\bf k}_{i\perp})
\phi_{M}(x_{i},{\bf k}_{i\perp}),
\eea
where ${\cal R}^{M}_{\lambda_{q}\lambda_{\bar q}}$ is the spin-orbit 
wavefunction, which is obtained by the interaction independent 
Melosh transformation from the ordinary equal-time static spin-orbit
wavefunction, and $\phi_{M}(x_{i},{\bf k}_{i\perp})$ is the radial 
wavefunction.
The operators $\Gamma_{M}$ for the scalar(S) and vector(V) mesons are
given by
\bea{\label{eq:3}}
\Gamma_{S}&=& \frac{({\not\!\!\!\:P_S}+M^S_0)
\bigg(\frac{K\cdot P_S}{M^S_0}-{\not\!\!\!\:K}\bigg)}
{(M^S_0+m_q+m_{\bar{q}})\sqrt{2[(M_0^S)^2-(m_q^2-m^2_{\bar{q}})]}}\nonumber\\
\Gamma_{V}&=&\frac{-({\not\!\!\!\:P_V}+M^V_0)\not\!\epsilon(J_3)}
{(M^V_0+m_q+m_{\bar{q}})\sqrt{2[(M_0^V)^2-(m_q^2-m^2_{\bar{q}})]}},\:
\eea
where $P_{(S,V)}\equiv(p_q+p_{\bar{q}})$, $K\equiv(p_{\bar{q}}-p_q)/2$
is the relative four-momentum between the quark and antiquark, and
$\epsilon$ is the polarization four-vector of the vector meson(with
momentum $P_V$), which is given by
\bea{\label{pol_vec}}
\epsilon^\mu(\pm)&=&[\ep^+,\ep^-,\ep_\perp]
=\biggl[0,\frac{2}{P^+_V}{\bf\epsilon}_\perp(\pm)\cdot{\bf P_{V\perp}},
{\bf\epsilon}_\perp(\pm)\biggr],
\nonumber\\
{\bf\epsilon}_\perp(\pm)&=&\mp\frac{(1,\pm i)}{\sqrt{2}},
\nonumber\\
\epsilon^\mu(0)&=&
\frac{1}{M_V}\biggl[P^+_V,\frac{{\bf P}^2_{V\perp}-M^2_V}{P^+_V},
{\bf P}_{V\perp}\biggr].
\eea
The operator $\Gamma_V$ was derived in Ref.~\cite{JCC1992}. We followed the same
procedure, detailed in Ref.~\cite{JCC1992}, in order to derive $\Gamma_S$. Note that
in the case of the vector meson, the operator $\Gamma_V$ has the expected form,
${({\not\!\!\!\:P}+M)\not\!\epsilon}$.  However, because the scalar ($^3P_0$) state
possesses non-zero orbital angular momentum, the proper Melosh--transformed
spin-orbit wavefunction is not simply given by $\Gamma_S=({\not\!\!\!\:P}+M)$,
as one might expect.  The form is more complicated as shown in 
Eq.(\ref{eq:3}), and it depends explicitly upon the
relative momentum between the meson's constituents.  This same type of behavior
was demonstrated for the axial--vector meson in Ref.~\cite{JCC1992}.  Since the
axial--vector ($^3P_1$) state also possesses non-zero orbital angular momentum, the
spin-orbit wavefunction is not simply given by
${({\not\!\!\!\:P}+M)\not\!\epsilon\gamma^5}$.  The correct form contains an
additional factor which explicitly depends on the relative momentum between the
quark and anti-quark.  It is interesting to note,
however, that in the case where $m_q=m_{\bar{q}}=m$ (which we use throughout this
work), the expressions in Eq.~(\ref{eq:3}) reduce to
\bea{\label{eq:3.5}}
\Gamma_{S}&=& \frac{1}{2\sqrt{2}M_0^S}({\not\!\!\!\:P_S}+M_0^S),\nonumber\\
\Gamma_{V}&=&\frac{-1}{\sqrt{2}M_0^V(M^V_0+2m)}({\not\!\!\!\:P_V}+M^V_0)\not\!\epsilon(J_3).
\eea
So, in the equal mass case, $\Gamma_S$ does have the expected form.
We confirmed the similar reduction of the axial-vector meson wavefunction 
in the equal mass case. 
These expressions can be further simplified to the form we will use
in our analysis:
\bea{\label{eq:3.75}}
\Gamma_{S}&=& \frac{1}{2\sqrt{2}},\nonumber\\
\Gamma_{V}&=&\frac{-1}{\sqrt{2}M_0^V}\bigg[\not\!\epsilon(J_3)-
\frac{\epsilon\cdot(p_q-p_{\bar{q}})}{M^V_0+2m}\bigg].
\eea
The spin-orbit wavefunctions satisfy the following relations
\bea{\label{eq:4}}
\sum_{\lambda_{q}\lambda_{\bar q}}
{\cal R}^{S\dagger}_{\lambda_{q}\lambda_{\bar q}}
{\cal R}^{S}_{\lambda_{q}\lambda_{\bar q}} &=& 
\frac{1}{4}(M^{2}_{0} - 4m^{2})
= |{\bf k}|^{2}, \nonumber\\
\sum_{\lambda_{q}\lambda_{\bar q}}
{\cal R}^{V\dagger}_{\lambda_{q}\lambda_{\bar q}}
{\cal R}^{V}_{\lambda_{q}\lambda_{\bar q}} &=& 1,
\eea
where ${\bf k}=({\bf k}_{\perp},k_{z})$ is the three momentum of the
constituent quark and $k_{z}= (x-\frac{1}{2})M_{0}$. 
Note that the total wavefunction $\Psi_{S}(x,{\bf k}_{\perp})$
for the scalar meson vanishes at $|{\bf k}|=0$ in accordance
with the property of $P$-wave function.

For the radial wavefunctions $\phi_M(x_i,{\bf k}_{i\perp})$,
we shall use the following gaussian wavefunctions 
for the scalar and vector mesons 
\bea{\label{eq:5}}
\phi_S(x,{\bf k}_\perp)&=&
{\cal N}\sqrt{\frac{2}{3\beta^2}}
\sqrt{\frac{\partial k_z}{\partial x}}
\exp(-{\bf k}^2/2\beta^2),
\nonumber\\
\phi_V(x,{\bf k}_\perp)&=& {\cal N}
\sqrt{\frac{\partial k_z}{\partial x}}
\exp(-{\bf k}^2/2\beta^2),
\eea
where ${\cal N}= 4({\pi\over\beta^2})^{3/4}$ 
and ${\partial k_{z}}/{\partial x}$ is the Jacobian of
the variable transformation $\{x,{\bf k}_{\perp}\}\to {\bf k}=
(k_{z},{\bf k}_{\perp})$ defined by
\bea\label{jacob}
\frac{\partial k_z}{\partial x}&=& \frac{M_{0}}{4x(1-x)},
\eea
and the normalization factors are obtained from the
following normalization of the total wavefunction,
\be
\int_0^1 dx \int {\frac{d^2{\bf k}_{\perp}}{16\pi^3}}
|\Psi_{M}(x_{i},{\bf k}_{\perp})|^2=1.
\ee

The wavefunctions depend on only two model parameters: the constituent quark
mass, $m$, and the binding strength, $\beta$.  For these parameters, we use the
values determined in Ref.~\cite{CJ1} which are ($m_{u,d}\equiv m_n=\text{0.22 GeV}$,
$\beta_{u,d}\equiv \beta_n=\text{0.3659 GeV}$) and
($m_s=\text{0.45 GeV}$,$\beta_s=\text{0.4128 GeV}$).  These values were obtained
by fitting the LFQM spectrum for pseudoscalar and vector mesons---obtained using
a QCD-inspired model Hamiltonian with a linear confining potential---to experimental
data.  In this paper, we will use these values of the parameters for both vector
and scalar mesons.  In a future work, we intend to perform a separate fit to
scalar meson data to see whether or not the model parameters would
differ significantly.

Since our analysis deals with $\phi$ decays, a value for the $\omega$--$\phi$
mixing angle is necessary.  We use $\delta_{\omega\text{-}\phi}=\pm7.8\,^{\circ}$
(the sign cannot be fixed), which was also determined in Ref.~\cite{CJ1}.  This
value was obtained by a mass squared mixing analysis in which it was assumed that
\bea
|\phi\ra&=&-\sin\delta_{\omega\text{-}\phi}|n\bar{n}\ra - \cos\delta_{\omega\text{-}\phi}|s\bar{s}\ra \nonumber \\
|\omega\ra&=&\;\;\;\cos\delta_{\omega\text{-}\phi}|n\bar{n}\ra - \,\sin\delta_{\omega\text{-}\phi}|s\bar{s}\ra, \nonumber
\eea
and in which the masses of the bare quarkonia were determined using the
model parameters and Hamiltonian mentioned in the previous paragraph.
\section{Scalar Mixing Amplitudes}
Glueball--$q\bar{q}$ mixing can be described using a mass mixing matrix.
Written in the $|gg\ra$, $|s{\bar s}\ra$, $|n{\bar n}\ra$ basis, this
takes the form \cite{LW1999,CK2001}
\be
M=\left( \begin{array}{ccc}
M_{gg}& f& \sqrt{2}fr\\
f  & M_{s{\bar s}}& 0\\
\sqrt{2}fr& 0& M_{n{\bar n}}
\end{array} \right) \: ,
\ee 
where $f=\la gg|M|s{\bar s}\ra$ and $r=\la gg|M|n{\bar n}\ra 
/\sqrt{2}\la gg|M|s{\bar s}\ra$.
As described in the introduction, the mixing is assumed
to be among the three isoscalar states $f_0(1370),f_0(1500)$ and
$f_0(1710)$.  These mixed physical states can be written as
\bea\label{mix1}
\left(
\begin{array}{c}
|f_0(1710)\ra \\ |f_0(1500)\ra \\ |f_0(1370)\ra
\end{array}\right)\;&=&
\left(
\begin{array}{ccc}
a_{1}& b_{1}& c_{1}\\
a_{2}& b_{2}& c_{2}\\
a_{3}& b_{3}& c_{3}
\end{array}\right)\;
\left(
\begin{array}{c}
|gg\ra \\ |s\bar{s}\ra \\ |n{\bar n}\ra
\end{array}\right), 
\eea
The mixing amplitudes, $a_i$, $b_i$, and $c_i$ can be written in terms
of the physical masses ($M_{f_j}$), the bare masses ($M_{gg}$,
$M_{s{\bar s}}$, and $M_{n{\bar n}}$), and the glueball--bare
quarkonia mixing strengths ($f$ and $r$).  
For the present analysis we will adopt the values obtained by Lee and 
Weingarten, and by Close and Kirk.

Beginning with the lattice QCD--motivated assumption that the bare
scalar $s\bar{s}$ is lighter than the scalar glueball
(\emph{i.e.} $M_{n{\bar n}}<M_{s{\bar s}}<M_{gg}$), Lee and Weingarten
obtained the following mixing amplitudes \cite{LW1999}:
\be \label{Weingarten}
\left(
\begin{array}{ccc}
\hspace{0.25cm}0.86\pm0.05&\hspace{0.2cm}0.30\pm0.05&\hspace{0.25cm}0.41\pm0.09\\
-0.13\pm0.05&\hspace{0.2cm}0.91\pm0.04&-0.40\pm0.11\\
-0.50\pm0.12&\hspace{0.2cm}0.29\pm0.09&\hspace{0.25cm}0.82\pm0.09
\end{array}\;\;\right)\;. 
\ee
Using a different approach, Close and Kirk \cite{CK2001} examined the constraints
placed on the flavor content of $f_0(1370)$, $f_0(1500)$, and $f_0(1710)$ by
decay branching ratios to pairs of pseudoscalar mesons.
From a $\chi^2$ analysis of the available branching ratio data,
they obtained various solutions depending on which parameters were left free
at the outset.  One solution was consistent with a glueball which
lies just above the bare $n{\bar n}$ (\emph{i.e.} $M_{n{\bar n}}
\lesssim M_{gg}<M_{s{\bar s}}$).  It's associated mixing amplitudes are 
\be \label{Close}
\left(
\begin{array}{ccc}
\hspace{0.25cm}0.39\pm0.03&\hspace{0.2cm}0.91\pm0.02&\hspace{0.25cm}0.15\pm0.02\\
-0.65\pm0.04&\hspace{0.2cm}0.33\pm0.04& -0.70\pm0.07\\
-0.69\pm0.07&\hspace{0.2cm}0.15\pm0.01&\hspace{0.25cm}0.70\pm0.07
\end{array}\;\;\right)\;. 
\ee
Another of their solutions was consistent with an even lighter glueball which
lies below the mass of the $n\bar{n}$ (\emph{i.e.} $M_{gg}<M_{n\bar{n}}<M_{s\bar{s}}$).
The mixing amplitudes for this solution are
\be \label{Close2}
\left(
\begin{array}{ccc}
\hspace{0.4cm}0.25&\hspace{0.5cm}0.96&\hspace{0.5cm}0.10\\
\hspace{0.15cm}-0.37&\hspace{0.5cm}0.13&\hspace{0.25cm}-0.92\\
\hspace{0.15cm}-0.89&\hspace{0.5cm}0.14&\hspace{0.5cm}0.44
\end{array}\;\;\right)\;. 
\ee

In the next section, we shall obtain expressions for the transition
form factors for various radiative decays. Then, in the following 
section, we will use the
mixing amplitudes given in Eqs.~(\ref{Weingarten})--(\ref{Close2})
above to numerically evaluate the form factors and corresponding
decay widths for the heavy, medium, and light weight glueball
cases respectively.  
\section{Form factors for radiative decays}
\subsection{The process $V(S)\to S(V) + \gamma$}
The coupling constant $g_{AX\gamma}$ 
for the radiative $P_A(q_{1}\bar{q})\to P_X(q_{2}\bar{q})\gamma$
decays between vector (V) and scalar (S) mesons, {\it i.e.}
$(A,X)=(^{3}S_{1},^{3}P_{0})$ or $(^{3}P_{0},^{3}S_{1})$,
is obtained by the matrix element of the electromagnetic 
current $J^{\mu}$ which is defined by
\bea{\label{eq:M_VS}}
{\cal M}_{1}&=&\la X(P_2)|\epsilon_{\gamma}\cdot J|A(P_1)\ra
\nonumber\\
&=&eg_{AX\gamma}[(\epsilon_{\gamma}\cdot\epsilon_{V})(P_1\cdot q)
- (\epsilon_{\gamma}\cdot P_1)(\epsilon_{V}\cdot q)],\nonumber
\\
\eea
where $\epsilon_{\gamma}$ and $\epsilon_{V}$
are the polarization vectors of the photon and the vector meson,
respectively. Since the $J_{z}=0$ state of the vector meson cannot
convert into a real photon, the $\epsilon_{V}$ should be transversely
polarized($J_{z}=\pm 1$) to extract the coupling constant
$g_{AX\gamma}$. In other words, the possible helicity combinations
in the transition $V(S)\to S(V)\gamma$ are either from ($J^{\gamma}_{z}=+1,
J^{V}_{z}=-1$) or from ($J^{\gamma}_{z}=-1,J^{V}_{z}=+1$).
The decay width for $A\to X + \gamma$ is given by \cite{choi2}
\bea{\label{eq:D_VS}}
\Gamma(A\to X + \gamma)
&=&\alpha\frac{g^{2}_{AX\gamma}}{2J_{A} +1}
\biggl[\frac{M^{2}_{A} - M^{2}_{X}}{2M_{A}}\biggr]^{3},
\eea
where $\alpha$ is the fine structure constant and $J_{A}$ is the total
angular momentum of the initial particle.

\begin{figure}[t]
\includegraphics[width=2.75in]{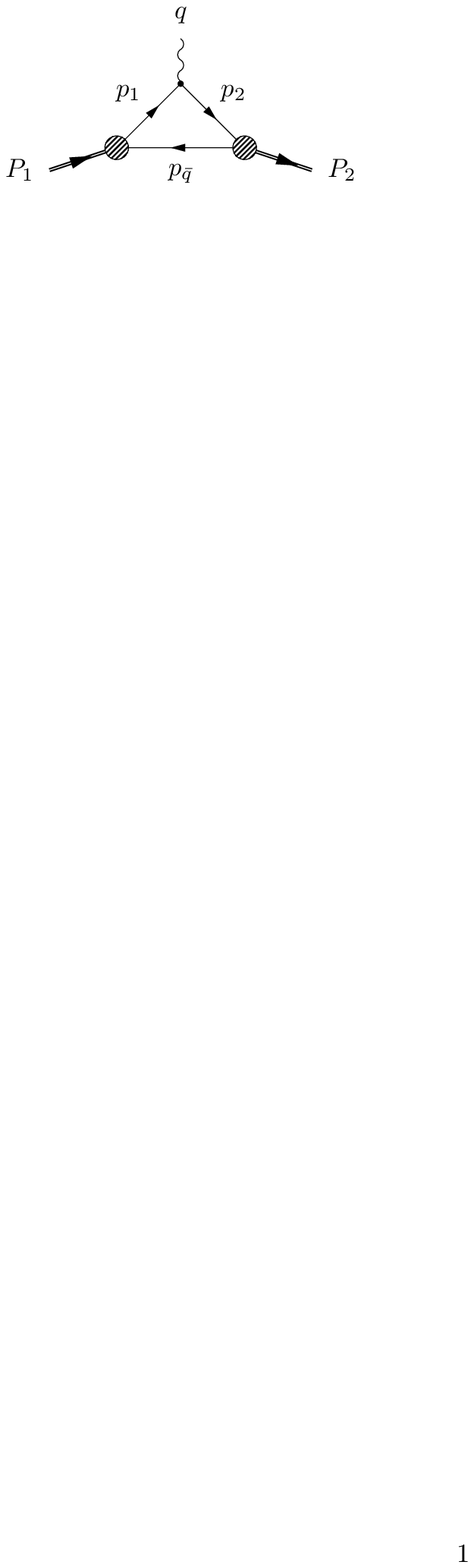}
\caption{Primary diagram for $A(P_1) \to X(P_2) +\gamma(q)$.  There is an additional
diagram in which the virtual photon interacts with the antiquark.}
\label{fig:sv}
\end{figure}

In LFQM calculations, we shall analyze the virtual photon($\gamma^*$)
decay process so that we calculate the momentum dependent transition
form factor, $F_{AX\gamma^*}(Q^2)$. The coupling constant, $g_{AX\gamma}$,
can then be determined in the limit as $Q^2\to 0$ ({\it i.e.},
$g_{AX\gamma}=F_{AX\gamma^*}(Q^2=0)$).
Figure~\ref{fig:sv} shows the primary Feynman diagram for this process.
The amplitude is
\bea\label{form}
{\cal M}^{\mu}_1&=&\la X(P_2)|J^{\mu}|A(P_1)\ra
\nonumber\\
&=& eF_{AX\gamma}(q^2)[\epsilon_{V}^{\mu}(P_1\cdot q)
- P_1^{\mu}(\epsilon_{V}\cdot q)].
\eea
 Our analysis is based on the standard light-front 
frame($q^{+}=0$)\cite{Brod}.
\bea{\label{eq:DY}}
P_1&=& (P^+_1,P^{-}_1,{\bf P}_{1\perp})= 
(P^{+}_1, \frac{M_{A}^{2}}{P^{+}_1},
{\bf 0}_{\perp}), \nonumber\\
q&=& (0, \frac{M^{2}_{A} - M^{2}_{X} - Q^{2}}{P^{+}},{\bf q}_{\perp}),
\nonumber\\
P_2&=& P_1 - q = (P^{+}_1,\frac{M^{2}_{X} + Q^{2}}{P^{+}_1}, -{\bf q}_{\perp}),
\eea
where ${\bf q}^{2}_{\perp}=Q^{2}= -q^{2}$ is the space-like photon momentum
transfer.

The quark momentum variables in the $q^{+}=0$ frame are given by
\bea\label{vari}
p^{+}_{1}&=&(1-x)P^{+}_{1},\;
p^{+}_{\bar{q}}=xP^{+}_{1},\nonumber\\
{\bf p}_{1\perp}&=&(1-x){\bf P}_{1\perp} + {\bf k}_{\perp},\;
{\bf p}_{\bar{q}\perp}= x{\bf P}_{1\perp} - {\bf k}_{\perp},
\nonumber\\
p^{+}_{2}&=&(1-x)P^{+}_{2},\;
p'^{+}_{\bar{q}}=xP^{+}_{2},\nonumber\\
{\bf p}_{2\perp}&=&(1-x){\bf P}_{2\perp} +
{\bf k'}_{\perp}, \;
{\bf p'}_{\bar{q}\perp}= x{\bf P}_{2\perp} - {\bf k'}_{\perp},
\eea
which require that $p^{+}_{\bar{q}}=p'^{+}_{\bar{q}}$
and ${\bf p}_{\bar{q}\perp}={\bf p'}_{\bar{q}\perp}$.

The hadronic matrix elements in the radiative decay between
two particles $A$ and $X$ are of the form
\bea\label{Conv}
&&\la X(P_2)|J^\mu|A(P_1)\ra
\nonumber\\
&&=\sum_{j,{\lambda}'s}{\cal A}_j\int^1_0 dx\int {\frac{d^2{\bf k}_\perp}
{16\pi^3}}
\phi_{2}(x,{\bf k'}_\perp)\phi_{1}(x, {\bf k}_\perp)
\nonumber\\
&&\times
{\cal R}^{X^\dagger}_{\lambda_2\bar{\lambda}}
\frac{\bar{u}_{\lambda_2}(p_2)}{\sqrt{p^+_2}}\gamma^\mu
\frac{u_{\lambda_1}(p_1)}{\sqrt{p^+_1}}
{\cal R}^{A}_{\lambda_1\bar{\lambda}}
\nonumber\\
&&=e\sum_{j}{\cal A}_jI_1(m_j,Q^2)[\epsilon_{V}^{\mu}(P_1\cdot q)
- P_1^{\mu}(\epsilon_{V}\cdot q)],
\eea
where ${\bf k'}_\perp={\bf k}_\perp - x{\bf q}_\perp$, ${\cal A}_j$ is the
overlap of the $j^{th}$ flavor portion of the flavor wavefunctions and
contains all of the relevant charge factors and mixing amplitudes.  Comparing
the last line of Eq.~(\ref{Conv}) with Eq.~(\ref{form}) it can be seen that we have
defined the form factor in terms of the charge and mixing amplitude independent
quantity $I_1(m_j,Q^2)$. 

The sum of the light-front spinors over the helicities in Eq.~(\ref{Conv})
is obtained as
\bea\label{trace}
S^\mu&=&\sum_{\lambda's}{\cal R}^{X^\dagger}_{\lambda_2\bar{\lambda}}
\frac{\bar{u}_{\lambda_2}(p_2)}{\sqrt{p^+_2}}\gamma^\mu
\frac{u_{\lambda_1}(p_1)}{\sqrt{p^+_1}}
{\cal R}^{A}_{\lambda_1\bar{\lambda}}
\nonumber\\
&=&\frac{
{\rm Tr}[(\not\!p_{\bar q}-m_j)\Gamma_X
(\not\!p_2+m_j)\gamma^\mu(\not\!p_1+m_j)\Gamma_A]}
{\sqrt{p^+_1p^+_2}}. \nonumber
\\
\eea
The explicit form of the trace is summarized in the Appendix.
For $V(P_1)\to S(P_2)\gamma^*(q)$ decay, the (transverse) polarization 
vector $\epsilon_V$ of the vector meson is given by 
$\epsilon_V(\pm)=[0,0,\epsilon_\perp(\pm)]$ and the trace term 
with the plus component of the current is given by
\begin{widetext}
\bea\label{S_VS}
S^+_{V\to S}\hspace{-0.2cm}&=&\frac{-1}{(1-x)M_0\sqrt{2}}
\biggl\{ m_j[q^R-4(1-x)k^R]
-\frac{2k^R}{x(M_0+2m_j)}
[(2x-1)^2m_j^2 +{\bf k}^2_\perp -x{\bf k}_\perp\cdot{\bf q}_\perp]
\biggr\},
\eea
where we use $\epsilon_V(+)$.  Therefore, we obtain the one-loop integral,
$I_1(m_j,Q^2)$, as follows
\bea{\label{g_VS}}
I_1(m_j,Q^2)
&=& 
\int^1_0 dx\int {\frac{d^{2}{\bf k}_{\perp}}{16\pi^3}}
\phi_S(x,{\bf k'}_\perp)\phi_V(x,{\bf k}_\perp)
\frac{1}{x(1-x)M_0} \biggl\{ xm_j[1-4(1-x)\frac{k^R q^L}{{\bf q}^2_\perp}]
\nonumber\\
&&-\frac{2(k^R q^L)/{\bf q}^2_\perp}{(M_0+2m_j)}
[(2x-1)^2m_j^2 +{\bf k}^2_\perp -x{\bf k}_\perp\cdot{\bf q}_\perp]
+ (x\to 1-x,{\bf k}_\perp\to -{\bf k}_\perp)
\biggr\},
\eea
\end{widetext}
where $k^R q^L={\bf k}_\perp\cdot{\bf q}_\perp 
- i|{\bf k}_\perp\times{\bf q}_\perp|_z$ and even though the cross
term does not contribute to the integral, the dot product term does
contribute.

Then, the transition form factor $F_{VS\gamma^*}(Q^2)$ is given by
\bea\label{F_VS1}
F_{VS\gamma^*}(Q^2)=
{\cal A}_nI_1(m_n,Q^2)+{\cal A}_sI_1(m_s,Q^2),
\eea
where ${\cal A}_n$ and ${\cal A}_s$ are the overlaps of the up--down and
strange portions of the flavor wavefunctions respectively. For example,
in the case where $V=\phi$ and $S=(u\bar{u}-d\bar{d})/\sqrt{2}$,
${\cal A}_n=-(\sin\delta_{\omega\text{--}\phi})(e_u-e_d)/2$ and 
${\cal A}_s=0$.  
Also, for $V=\phi$ and $S=s\bar{s}$, ${\cal A}_n=0$ and
${\cal A}_s=-(\cos\delta_{\omega\text{--}\phi})e_s$.  

The transition form factor $F_{SV\gamma^*}(Q^2)$ for 
$S(P_1)\to V(P_2)\gamma^*(q)$ can be obtained from Eq.~(\ref{g_VS})
by replacing ${\bf q}_\perp\to -{\bf q}_\perp$ and
${\bf k}_\perp\to {\bf k'}_\perp$ and the explicit form for the
one-loop integral corresponding to Eq.~(\ref{g_VS}) is given by 
\begin{widetext}
\bea{\label{g_SV}}
I'_1(m_j,Q^2)
&=&
\int^1_0 dx\int {\frac{d^{2}{\bf k}_{\perp}}{16\pi^3}}
\phi_V(x,{\bf k'}_\perp)\phi_S(x,{\bf k}_\perp)
\frac{1}{x(1-x)M'_0} 
\biggl\{ xm_j[-(2x-1)^2-4(1-x)\frac{k^R q^L}{{\bf q}^2_\perp}]
\nonumber\\
&&-\frac{2(k^R q^L-x{\bf q}^2_\perp)/{\bf q}^2_\perp}{(M'_0+2m_j)}
[(2x-1)^2m_j^2 +{\bf k}^2_\perp -x{\bf k}_\perp\cdot{\bf q}_\perp]
+ (x\to 1-x; {\bf k}\to -{\bf k}_\perp)
\biggr\},
\eea
\end{widetext}
where $M'^2_0=(m_j^2+{\bf k'}^2_\perp)/x(1-x)$ and again the cross
term in $k^R q^L$ does not contribute to the integral.
As one may expect, however, we found that $I'_1(m_j,Q^2)=-I_1(m_j,Q^2)$ and thus
$F_{SV\gamma^*}(Q^2)=-F_{VS\gamma^*}(Q^2)$.

\subsection{The process $S\to\gamma\gamma$}
We now apply this model calculation to the two photon decays of scalar 
mesons.   
In this case, the coupling constant $g_{S\gamma\gamma}$ for $S\to\gamma\gamma$ 
is defined by
 
\bea{\label{eq:28}}
{\cal M}_{2}&=&\la \gamma(P_2)|\epsilon_{1}\cdot J|S(P_1)\ra
\nonumber\\
&=& e^{2}g_{S\gamma\gamma}[(\epsilon_{1}\cdot\epsilon_{2})(P_1\cdot q) -
(\epsilon_{1}\cdot P_1)(\epsilon_{2}\cdot q)],   
\eea
\\
where $\epsilon_{1}=\epsilon_{\gamma}(q)$ and 
$\epsilon_{2}=\epsilon_{\gamma}(P_2)$. In terms of $g_{S\gamma\gamma}$,
the decay width for this process is given by
\begin{eqnarray}{\label{eq:33}}
\Gamma = \frac{\pi}{4}\alpha^{2}g^{2}_{S\gamma\gamma}M^{3}_{S}\:.
\end{eqnarray}
Here again, instead of calculating the two real photon decays,
we first calculate the matrix element for $S\to\gamma\gamma^*$,
which is given by 
\bea\label{Sgg*}
{\cal M}_2^{\mu}&=& \la \gamma(P_2)|J^\mu|S(P_1)\ra,\nonumber\\
&=&e^2F_{S\gamma\gamma^*}(Q^2)[\epsilon_2^{\mu}(P_1\cdot q)-P_1^{\mu}(\epsilon_2\cdot q)],
\eea 
and take the limit $Q^2\to 0$ to compute the decay rate for the
two real photon decays. Using the same quark momentum variables in $q^{+}=0$ frame 
as Eq.~(\ref{vari}) with the plus component of the current, we then obtain  
\begin{widetext}
\begin{eqnarray}{\label{Gamp}}
{\cal M}_2^+&=& \sqrt{n_{c}} \sum_j {\cal A}_j 
\sum_{\lambda_1,\lambda_2,{\bar\lambda}}\int^{1}_{0}dx
\int {\frac{d^{2}{\bf{k_{\perp}}}}{16\pi^3}} 
\phi_{S}(x_{i},{\bf{k_{\perp}}})
\biggl[\frac{\bar{v}_{\bar\lambda}(p_{\bar q})}{\sqrt{p^+_{\bar q}}}
\not\!\epsilon_{2}\frac{u_{\lambda_2}(p_2)}{\sqrt{p^+_2}}
\frac{\bar{u}_{\lambda_2}(p_2)}{\sqrt{p^+_2}}\gamma^+
\frac{u_{\lambda_1}(p_1)}{\sqrt{p^+_1}}
\nonumber\\
&&\;\times
\frac{1}{{\bf q}_{\perp}^{2}- [{\bf p}^2_{2\perp}
+ m_j^{2}]/p^+_2 - [{\bf p}^2_{{\bar q}\perp} + m_j^{2}]/p^+_{\bar q}} 
+ (x\to 1-x,{\bf k}_\perp\to -{\bf k}_\perp)\biggr]\;
{\cal R}^S_{\lambda_1\lambda_{\bar q}}\nonumber\\
&=&e^2\bigg(\sum_j {\cal A}_j I_2(m_j,Q^2)\bigg)
[\epsilon_2^{\mu}(P_1\cdot q)-P_1^{\mu}(\epsilon_2\cdot q)],
\end{eqnarray}
where ${\cal A}_j$ now contains factors of $e_j^2$ due to the presence
of two electromagnetic vertices. As in the case of Eq.~(\ref{trace}), the sum of
the light-front spinors over the helicities in the first term of Eq. ~(\ref{Gamp})
is obtained as
\bea{\label{trace2}}
T^\mu&=&\sum_{\lambda_1,\lambda_2,{\bar\lambda}}
\frac{\bar{v}_{\bar\lambda}(p_{\bar q})}{\sqrt{p^+_{\bar q}}}
\not\!\epsilon_{2}\frac{u_{\lambda_2}(p_2)}{\sqrt{p^+_2}}
\frac{\bar{u}_{\lambda_2}(p_2)}{\sqrt{p^+_2}}\gamma^+
\frac{u_{\lambda_1}(p_1)}{\sqrt{p^+_1}}
{\cal R}^S_{\lambda_1\lambda_{\bar q}}
\nonumber\\
&=&\frac{{\rm Tr}[(\not\!p_{\bar q}-m_j)\not\!\epsilon_2(\not\!p_2+m_j)
\gamma^+(\not\!p_1+m_j)]}{2\sqrt{2p^+_{\bar q}(p^+_2)^2 p^+_1}}
\nonumber\\
&=&\frac{4m_j[p^+_1(p_{\bar q}\cdot\epsilon_2-p_2\cdot\epsilon_2)
+p^+_2(p_{\bar q}\cdot\epsilon_2-p_1\cdot\epsilon_2)
+p^+_{\bar q}(p_2\cdot\epsilon_2-p_1\cdot\epsilon_2)]}
{2\sqrt{2p^+_{\bar q}(p^+_2)^2 p^+_1}},
\eea
where we have used the fact that $\epsilon^+_2(\pm)=0$.
Now, using $\epsilon_2(+)=(0,\sqrt{2}q^R/P^+_1,{\bf\epsilon}_\perp)$,
we finally obtain the one loop integral,
\bea{\label{Gamp2}}
I_2(m_j,Q^2)
&=&-\sqrt{6} \int^1_0 dx\int {\frac{d^2{\bf k}_\perp}{16\pi^3}}
\phi_S(x,{\bf k}_\perp)
\biggl\{
\frac{m_j[(2x-1)^2 
+ 4(1-x)(k^R q^L/{\bf q}^2_\perp)]}{(1-x)\sqrt{x(1-x)}}
\frac{x(1-x)}{m_j^2 + ({\bf k}_\perp-x{\bf q}_\perp)^2}
\nonumber\\
&&+(x\to 1-x,{\bf k}_\perp\to -{\bf k}_\perp)
\biggr\},
\eea
\end{widetext}
and the transition form form factor is given by
\bea\label{Fsgg}
F_{S\gamma\gamma^*}(Q^2)
={\cal A}_n I_2(m_n,Q^2)
+{\cal A}_s I_2(m_s,Q^2).
\eea  
As an example of the coefficients ${\cal A}_{n,s}$, consider the case
where $S=f_0(1370)$.  Here, ${\cal A}_n=c_3[(e_u^2+e_d^2)/\sqrt{2}]$
and ${\cal A}_s=b_3e_s^2$, where $c_3$ and $b_3$ are the glueball--quarkonia
mixing amplitudes of Eq.~(\ref{mix1}).

\section{Numerical results}
\subsection{Decays involving $f_0(1370)$, $f_0(1500)$, and $f_0(1710)$}
\begin{figure}[h]
\includegraphics[width=3.5in]{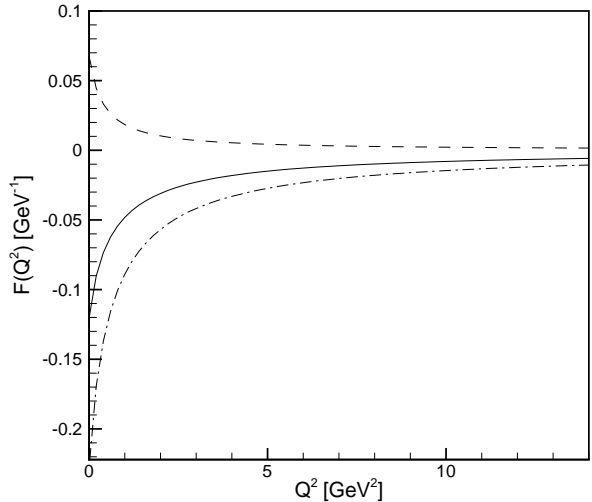}
\caption{$f_0\to\gamma\gamma^*$ transition form factors
for $f_0(1370)$ [dash-dotted], $f_0(1500)$ [dashed], and $f_0(1710)$ [solid].}
\label{fig:2photon}
\end{figure}

\begin{figure}
\includegraphics[width=3.5in]{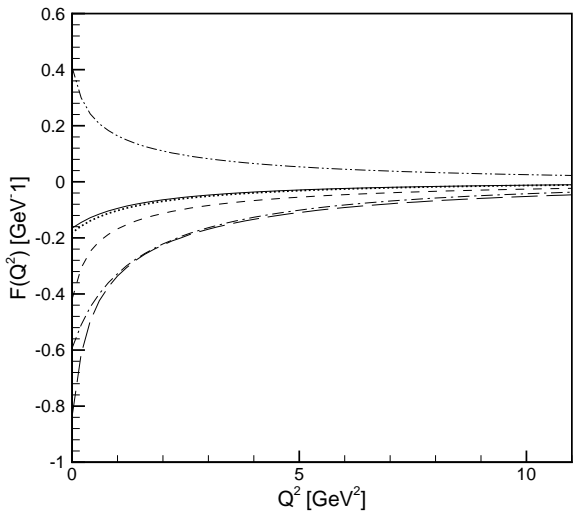}
\caption{$f_0\to\rho\gamma^*$ transition form factors for $f_0(1370)$ [long-dashed],
$f_0(1500)$ [dash-dot-dotted], and $f_0(1710)$ [short-dashed]; $f_0\to\phi\gamma^*$
transition form factors for $f_0(1370)$ [solid], $f_0(1500)$ [dash-dotted], and
$f_0(1710)$ [dotted].  Here we have used $\delta_{\omega\text{--}\phi}=
+7.8\,^{\circ}$.}
\label{fig:vector}
\end{figure}

\begin{figure}
\includegraphics[width=3.5in]{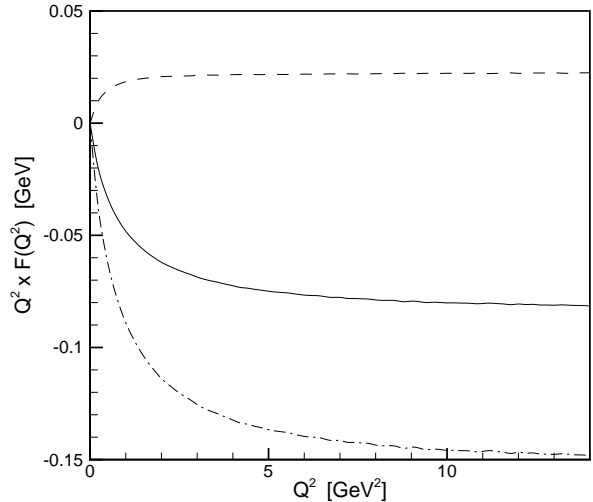}
\caption{$Q^2$ times the $f_0\to\gamma\gamma^*$ transition form factors
(Fig.~\ref{fig:2photon}) for $f_0(1370)$ [dash-dotted], $f_0(1500)$ [dashed],
and $f_0(1710)$ [solid].}
\label{fig:Qsq2photon}
\end{figure}

The expressions for the one-loop integrals, $I_1(m_j,Q^2)$ and $I_2(m_j,Q^2)$,
are evaluated numerically and used in Eqs.~(\ref{F_VS1}) and (\ref{Fsgg}) to compute
the $Q^2$-dependent transition form factors for $\gamma\gamma^*$,
$\phi\gamma^*$, and $\rho\gamma^*$ decays of the scalar mesons.
As an example, we give the results for the case of the heavy glueball
({\it i.e.} $M_{n\bar{n}}<M_{s\bar{s}}<M_{gg}$).  The transition form
factors for the $\gamma\gamma^*$ process are shown in
Fig.~\ref{fig:2photon}, and those for the $\phi\gamma^*$ and $\rho\gamma^*$
processes are collectively shown in Fig.~\ref{fig:vector}.  In the case of
the two photon decay, the form factor should fall off like $1/Q^2$
due to an intermediate quark propagator which becomes highly off--shell
at large $Q^2$.  Figure~\ref{fig:Qsq2photon} shows the behavior of
$Q^2\times F_{f_0\gamma\gamma^*}(Q^2)$ for each scalar state.  Each of the
curves clearly shows a tendency to flatten out, demonstrating $1/Q^2$
dependence in the form factors.

The decay constants for the real photon processes can be obtained from
the form factors in the limit as $Q^2\to0$ ({\it i.e.} $g=F(Q^2=0)$).  In
this limit, the values of the one-loop integrals are
\bea
I_1(m_n,Q^2=0)&=&2.05\:GeV^{-1}\nonumber\\
I_1(m_s,Q^2=0)&=&1.93\:GeV^{-1}\nonumber\\
I_2(m_n,Q^2=0)&=&-0.672\:GeV^{-1}\nonumber\\
I_2(m_s,Q^2=0)&=&-0.375\:GeV^{-1}.
\label{eq:oneloop}
\eea
The decay constants are obtained by substituting these values into Eqs.~(\ref{F_VS1})
and (\ref{Fsgg}).
The decay widths are then calculated using Eqs.~(\ref{eq:D_VS}) and (\ref{eq:33}).
The widths for the $\gamma\gamma$, $\phi\gamma$, and $\rho\gamma$ decays for
all of the glueball mass scenarios are summarized in Tables \ref{tab:photon},
\ref{tab:phi}, and \ref{tab:rho} respectively.  The uncertainties in these values
result solely from the uncertainties in the mixing amplitudes in Eqs.~(\ref{Weingarten})
and (\ref{Close}).  We have not accounted for the uncertainties in the meson masses.
The uncertainties in the masses of $f_0(1500)$ and $f_0(1710)$ ($\sim$0.4\%) are very
small compared to the uncertainties in the mixing amplitudes (6\%--40\%), and can
be neglected.  However, the uncertainty in the mass of $f_0(1370)$ is about 10\%,
and would, therefore, contribute significantly to the uncertainties in the decay
widths.  

Experimental data for radiative decays of the isoscalars
$f_0(1370)$, $f_0(1500)$, and $f_0(1710)$ are poor.  As one example, in the
recent past the PDG had reported partial widths of $3.8\pm1.5\:keV$ and
$5.4\pm2.3\:keV$ for the process $f_0(1370)\to\gamma\gamma$ \cite{groom}.
The PDG currently attributes these two values to $f_0(600)$, but at the same
time they state in a footnote that this data could equally well be assigned to
$f_0(1370)$ \cite{data}.  If these data which are on the order of a few keV do
belong to $f_0(1370)$, this would be encouraging given that our results listed
in Table I are consistent with this order of magnitude.  However, the ambiguity
noted above makes any such comparison irrelevant, and a great deal
more experimental investigation is necessary before any definitive conclusions can
be reached about the validity of any of the glueball mixing schemes.

In the absence of good experimental data with which to compare our results,
we turn to other theoretical predictions concerning these decay processes. 
In Ref.~\cite{CK2001}, Close and Kirk give predictions for ratios
of $f_0\to\gamma\gamma$ widths which depend only on charge factors
and mixing amplitudes, and ignore all mass-dependent effects.  For the ratios
$\Gamma(f_0(1710)\to\gamma\gamma)$:$\Gamma(f_0(1500)\to\gamma\gamma)$:$\Gamma
(f_0(1370)\to\gamma\gamma)$ they obtain
\bea
\text{Light Glueball}&=&1:5.1:2.8\nonumber\\
\text{Medium Glueball}&=&1:2.4:3.6\nonumber\\
\text{Heavy Glueball}&=&1:0.1:3.7 \:\:.
\eea
Our analysis which includes all of the relevant mass dependent effects yields
\bea
\text{Light Glueball}&=&1:8.7:1.7\nonumber\\
\text{Medium Glueball}&=&1:3.2:3.0\nonumber\\
\text{Heavy Glueball}&=&1:0.2:1.9 \:\:.
\eea
Our results differ slightly from those of Close and Kirk, however the
same overall qualitative pattern is preserved.

In Ref.~\cite{CDK2002}, Close, Donnachie, and Kalashnikova (CDK) compute
the $\phi$ and $\rho$ radiative decay widths for $f_0(1370)$, $f_0(1500)$,
and $f_0(1710)$ in the NR quark model.  Assuming that
$\phi=s\bar{s}$, they obtain the values listed in Tables \ref{tab:phiClose}
and \ref{tab:rhoClose}. Comparing these values with Tables \ref{tab:phi} and
\ref{tab:rho}, it is clear that the relativistic corrections introduced
by our model reduce the overall magnitudes of the decay widths by about
$50\text{--}70\%$. We note somewhat greater reduction for the process of
$f_0(1370) \to \phi\gamma$ due to our non-zero $\delta_{\omega{\text 
-}\phi}$.
A reduction in the widths would be expected given that the
relativistic motion of the constituents tends to spread out the meson's
wavefunction, thereby decreasing its peak value.  Despite the differences
in the overall magnitudes, however, the relative strengths between the different
decay processes are fairly well preserved.  Just as in CDK's analysis, we find
that the largest branching ratio is likely to be that of $f_0(1500)\to\rho\gamma$.
In our model, this branching ratio is about $1\%$ for the light glueball case,
$0.6\%$ for the medium glueball case, and $0.2\%$ for the heavy glueball case. 

\begin{table}
\caption{Decay widths for the process $f_0\to\gamma\gamma$.  The unit of the decay
width is [keV].  The uncertainties result from the uncertainties in the mixing
amplitudes in Eqs.~(\ref{Weingarten}) and (\ref{Close}).}
\begin{tabular}{|c|c|c|c|}
\hline
\vspace{-0.2cm}
&&&\\
&$\,$Light$\,$&$\,$Medium$\,$&Heavy\\
\hline
\vspace{-0.2cm}
&&&\\
$f_{0}(1370)$&
$1.6$&
$3.9^{+0.8}_{-0.7}$&
$5.6^{+1.4}_{-1.3}$\\[2pt]
\hline
\vspace{-0.2cm}
&&&\\
$f_{0}(1500)$&
$8.0$&
$4.1^{+1.0}_{-0.9}$&
$\,0.65^{+0.72}_{-0.45}$\\[2pt]
\hline
\vspace{-0.2cm}
&&&\\
$f_{0}(1710)$&
$0.92$&
$1.3^{+0.2}_{-0.2}$&
$3.0^{+1.4}_{-1.2}$\\[2pt]
\hline
\end{tabular}
\label{tab:photon}
\end{table}

\begin{table}
\caption{Decay widths for the process $f_0\to\phi\gamma$.
The unit of the decay width is [keV].  These values are for
$\delta_{\omega\text{-}\phi}=+7.8\,^{\circ}$. Using 
$\delta_{\omega\text{-}\phi}=-7.8\,^{\circ}$
does not significantly alter the results. The uncertainties result
from the uncertainties in the mixing amplitudes in
Eqs.~(\ref{Weingarten}) and (\ref{Close}).}
\begin{tabular}{|c|c|c|c|}
\hline
\vspace{-0.2cm}
&&&\\
&\phantom{A}Light\phantom{A}&\phantom{A}Medium\phantom{A}&
\phantom{A}Heavy\phantom{A}\\
\hline
\vspace{-0.2cm}
&&&\\
$f_{0}(1370)$&
$0.98$&
$0.83^{+0.27}_{-0.23}$&
$4.5^{+4.5}_{-3.0}$\\[2pt]    
\hline
\vspace{-0.2cm}
&&&\\
$f_0(1500)$&
$7.5$&
$28^{+7}_{-6}$&
$170^{+20}_{-20}$\\[2pt]    
\hline
\vspace{-0.2cm}
&&&\\
$f_0(1710)$&
$450$&
$400^{+20}_{-20}$&
$36^{+17}_{-14}$\\[2pt]
\hline   
\end{tabular}
\label{tab:phi}  
\end{table}

\begin{table}
\caption{Decay widths for the process $f_0\to\rho\gamma$.
The unit of the decay width is [keV]. The uncertainties result
from the uncertainties in the mixing amplitudes in
Eqs.~(\ref{Weingarten}) and (\ref{Close}).}
\begin{tabular}{|c|c|c|c|}
\hline
\vspace{-0.2cm}
&&&\\
&\phantom{A}Light\phantom{A}&\phantom{A}Medium\phantom{A}&
\phantom{A}Heavy\phantom{A}\\
\hline
\vspace{-0.2cm}
&&&\\
$f_{0}(1370)$&
$150$&
$390^{+80}_{-70}$&
$530^{+120}_{-110}$\\[2pt]  
\hline
\vspace{-0.2cm}
&&&\\
$f_{0}(1500)$&
$1100$&
$630^{+130}_{-120}$&
$210^{+130}_{-100}$\\[2pt]
\hline
\vspace{-0.2cm}
&&&\\
$f_{0}(1710)$&
$24$&
$55^{+16}_{-14}$&
$410^{+200}_{-160}$\\[2pt]
\hline
\end{tabular}
\label{tab:rho}
\end{table}

\begin{table}
\caption{CDK's results for $f_0\to\phi\gamma$.
The unit of the decay width is [keV].}
\begin{tabular}{|c|c|c|c|}
\hline
\vspace{-0.2cm}
&&&\\
&\phantom{A}Light\phantom{A}&\phantom{A}Medium\phantom{A}&
\phantom{A}Heavy\phantom{A}\\
\hline
\vspace{-0.2cm}
&&&\\
$f_{0}(1370)$&
$8$&
$9$&
$32$\\[2pt]    
\hline
\vspace{-0.2cm}
&&&\\
$f_0(1500)$&
$9$&
$60$&
$454$\\[2pt]    
\hline
\vspace{-0.2cm}
&&&\\
$f_0(1710)$&
$800$&
$718$&
$78$\\[2pt]
\hline   
\end{tabular}
\label{tab:phiClose}  
\end{table}

\begin{table}
\caption{CDK's results for $f_0\to\rho\gamma$.
The unit of the decay width is [keV].}
\begin{tabular}{|c|c|c|c|}
\hline
\vspace{-0.2cm}
&&&\\
&\phantom{A}Light\phantom{A}&\phantom{A}Medium\phantom{A}&
\phantom{A}Heavy\phantom{A}\\
\hline
\vspace{-0.2cm}
&&&\\
$f_{0}(1370)$&
$443$&
$1121$&
$1540$\\[2pt]  
\hline
\vspace{-0.2cm}
&&&\\
$f_{0}(1500)$&
$2519$&
$1458$&
$476$\\[2pt]
\hline
\vspace{-0.2cm}
&&&\\
$f_{0}(1710)$&
$42$&
$94$&
$705$\\[2pt]
\hline
\end{tabular}
\label{tab:rhoClose}
\end{table}

\subsection{Decays involving $a_0(980)$ and $f_0(980)$}
If we assume $a_0(980)$ to be a conventional $q\bar{q}$, then the flavor
structure should be $(u\bar{u}-d\bar{d})/\sqrt{2}$.  For the processes
$a_0(980)\to\gamma\gamma$ and $\phi\to a_0(980)\gamma$, the decay constants
and associated widths are calculated to be
\bea
g_{\phi a_0 \gamma}=-0.14\,GeV^{-1},&\Gamma_{\phi a_0 \gamma}=2.8\,eV\nonumber\\
g_{a_0 \gamma\gamma}=-0.16\,GeV^{-1},&\hspace{0.25cm}\Gamma_{a_0 \gamma\gamma}=990\,eV\:.
\eea
Our result for the magnitude of $g_{\phi a_0 \gamma}$ is consistent with 
the theoretical calculations
of Gokalp and Yilmaz \cite{GY2001}, who obtain $0.11\pm0.03\:GeV^{-1}$ 
using light-cone
QCD sum rules, and Titov {\it et al.} \cite{Titov1999}, who obtain 
$-0.16\:GeV^{-1}$ from
phenomenological considerations.  However, none of these calculated 
values for the
widths are consistent with experimental data.  The $\phi$ radiative width of
2.8 eV gives a branching ratio of $BR(\phi\to a_0\gamma)=6.7\times10^{-7}$ which
is significantly smaller than the PDG average of $0.88^{+0.17}_{-0.17}\times10^{-4}$;
and, the two-photon width of 990 eV is roughly 3 times larger than the value
reported by Amsler of $0.30\pm0.10\:keV$ \cite{Amsler98}.
The flavor content of the isoscalar $f_0(980)$ is less clear.  If we consider
the two possible extremes, $f_0(980)=n\bar{n}$ and $f_0(980)=s\bar{s}$, we obtain
\bea
n\bar{n}&=&\begin{cases}
g_{\phi f_0 \gamma}=-0.05\,GeV^{-1},&\Gamma_{\phi f_0 \gamma}=0.37\,eV\\
g_{f_0 \gamma\gamma}=-0.26\,GeV^{-1},&\Gamma_{f_0 \gamma\gamma}=2.7\,keV
\end{cases}\nonumber\\
\nonumber\\
s\bar{s}&=&\begin{cases}
g_{\phi f_0 \gamma}=+0.64\,GeV^{-1},&\Gamma_{\phi f_0 \gamma}=60\,eV\\
g_{f_0 \gamma\gamma}=-0.04\,GeV^{-1},&\Gamma_{f_0 \gamma\gamma}=63\,eV.
\end{cases}
\eea 
The PDG average for the two photon width is
$\Gamma_{f_0\gamma\gamma}=0.39^{+0.10}_{-0.13}\:keV$. Since the $s\bar{s}$
result falls below this value and the $n\bar{n}$ result sits above it,
it would be possible for some mixed $n\bar{n}\text{--}s\bar{s}$ state to reproduce the
data.  Working out the mixing required for this, we find that $f_0(980)$ would
be about 6\% $n\bar{n}$ and 94\% $s\bar{s}$.  This alone would allow for $f_0(980)$
to be interpreted as a conventional $q\bar{q}$.  However, the $\phi$ radiative
widths we calculated lead to the branching ratios
$BR(\phi\to f_0\gamma)=8.7\times10^{-8}$ for
the $n\bar{n}$ and $BR(\phi\to f_0\gamma)=1.4\times10^{-5}$ for
the $s\bar{s}$.  Both of these values fall well below the PDG average of
$3.3^{+0.8}_{-0.5}\times10^{-4}$.  Also, if we compute the ratio
$BR(\phi\to f_0\gamma)/BR(\phi\to a_0\gamma)$ for our model predictions,
we get 0.13 for $f_0=n\bar{n}$ and  21 for $f_0=s\bar{s}$, while the
experimental ratio is around 4.  This again hints at the possibility of
a mixed $n\bar{n}\text{--}s\bar{s}$ being able to reproduce the data.
However, the mixing needed to reproduce this ratio requires that $f_0(980)$
be 87\% $n\bar{n}$ and 13\% $s\bar{s}$.  This is the exact opposite of the
mixing needed to reproduce the two photon width above.
Overall our results are clearly inconsistent with the current experimental
data on $a_0(980)$ and $f_0(980)$.  Therefore, we conclude that these states
are most likely not $q\bar{q}$.

\section{Summary and Outlook}

We have performed the first LFQM calculations involving scalar mesons.  First,
the $^3P_0$ light--front wavefunction was constructed.  It was shown that, in
general, the covariant operator used to obtain the spin--orbit wavefunction depends
explicitly on the relative momentum between the meson's constituents, and is,
therefore, more complicated than the naive form that is commonly used.  This
wavefunction was used to compute radiative decays involving $f_0(1370)$, $f_0(1500)$,
$f_0(1710)$, $f_0(980)$, and $a_0(980)$.  In the case of the three heavy isoscalars,
the effects of glueball--$q\bar{q}$ mixing were taken into account.  Specifically,
three different mixing schemes corresponding to a heavy, medium, and light glueball
were used.  The lack of good experimental data made it difficult to draw
any conclusions about which of the three mixing scenarios, if any, could be the
correct one.  We note, however, that we have improved upon the earlier
NR model predictions of Close {\it et al.} \cite{CDK2002}.  Relativistic
corrections introduced by the LFQM resulted in decay widths that were about 50--70\%
smaller than those obtained in the NR calculations.  Yet, very little
change was observed in the pattern of relative strengths which is apparently quite
robust.  For the calculations involving
$a_0(980)$ and $f_0(980)$, we assumed these states to be $q\bar{q}$.  In contrast
to the case of the heavy scalars, there does exist well--established data for these
light scalars. While one or two of the properties we caculated, when taken in
isolation, could be considered consistent with the data, it is clear that
our results as a whole do not match the data.  This lends further support to
the current idea that $f_0(980)$ and $a_0(980)$ are not $q\bar{q}$ states.

In a future work, we intend to refine our analysis and perform our own
glueball--$q\bar{q}$ mixing calculation using the LFQM.  In this current work,
the model parameters ($m$, $\beta$) used for the scalar ($^3P_0$) meson
wavefunction were obtained from a spectrum calculation fit to
S--wave ($^1S_0$ and $^3S_1$) meson data.  In order to improve the scalar
wavefunction, we will perform a separate spectrum analysis using a QCD--inspired
model Hamiltonian similar to that of Ref.~\cite{CJ1}, and fit the spectrum to P--wave
($^3P_0$, $^3P_1$, $^3P_2$, and $^1P_1$) meson data.  In addition to refining
the model parameters, this analysis will also give the masses of the bare
$n\bar{n}$ and $s\bar{s}$ P--wave quarkonia.  With these masses, we will be
able to perform a glueball--$q\bar{q}$ mixing analysis involving the isoscalars
$f_0(1370)$, $f_0(1500)$, and $f_0(1710)$.  These mixing amplitudes, obtained
using the LFQM, could then be compared with those of Lee--Weingarten and Close--Kirk.

\appendix
\section{Spinor structure for $V(S)\to S(V)\gamma^*$ transition}
In this appendix, we show the explict form of the trace 
given by Eq. ~(\ref{trace}).
For the $V\to S\gamma^*$ transition, the following two trace 
calculations are necessary
\bea{\label{ap1}}
S^+_{VS1}&=&{\rm Tr}[(\not\!p_{\bar q}-m)
(\not\!p_2+m)\gamma^+(\not\!p_1+m)\not\!\epsilon]
\nonumber\\
&=&4m[p^+_1(\ep\cdot p_{\bar q} -\ep\cdot p_2)
      + p^+_2(\ep\cdot p_{\bar q}-\ep\cdot p_1)
\nonumber\\
&&+
 p^+_{\bar q}(\ep\cdot p_1 -\ep\cdot p_2)
\nonumber\\
&&
+\ep^+(p_1\cdot p_2 + p_2\cdot p_{\bar q}-p_1\cdot p_{\bar q}-m^2)],
\nonumber\\
S^+_{SV2}&=&{\rm Tr}[(\not\!p_{\bar q}-m)(\not\!p_2+m)\gamma^+(\not\!p_1+m)]
\nonumber\\
&=&4[p^+_1(p_2\cdot p_{\bar q}-m^2)+p^+_2(p_1\cdot p_{\bar q}-m^2)
\nonumber\\
&&+p^+_{\bar q}(m^2-p_1\cdot p_2)],
\eea
to get
\begin{equation}\label{ap2}
S^+_{V\to S}=\frac{-1}{4(1-x)M_0}
\biggl[S^+_{VS1} -\frac{\ep\cdot(p_1-p_{\bar q})}{M_0 + 2m}S^+_{VS2}
\biggr],
\end{equation}
where $\ep=\ep(P_1)$ and we used the transverse polarizations in the 
calculation of the form factor and decay width.

On the other hand, for the $S\to V\gamma^*$ transitions, we have
\bea{\label{ap3}}
S^+_{SV1}&=&{\rm Tr}[(\not\!p_{\bar q}-m)\not\!\epp
(\not\!p_2+m)\gamma^+(\not\!p_1+m)]
\nonumber\\
&=&4m[p^+_1(\epp\cdot p_{\bar q} -\epp\cdot p_2)
      + p^+_2(\epp\cdot p_{\bar q}-\epp\cdot p_1)
\nonumber\\
&& - p^+_{\bar q}(\epp\cdot p_1 -\epp\cdot p_2)
\nonumber\\
&& +\ep'^{+}(p_1\cdot p_2 + p_1\cdot p_{\bar q}-p_2\cdot p_{\bar q}-m^2)],
\nonumber\\
S^+_{SV2}&=&{\rm Tr}[(\not\!p_{\bar q}-m)(\not\!p_2+m)\gamma^+(\not\!p_1+m)]
\nonumber\\
&=&4[p^+_1(p_2\cdot p_{\bar q}-m^2)+p^+_2(p_1\cdot p_{\bar q}-m^2)
\nonumber\\
&& +p^+_{\bar q}(m^2-p_1\cdot p_2)],
\eea
to get
\begin{equation}\label{ap4}
S^+_{S\to V}=\frac{-1}{4(1-x)M'_0}
\biggl[S^+_{SV1} -\frac{\epp\cdot(p_2-p_{\bar q})}{M'_0 + 2m}S^+_{SV2}
\biggr],
\end{equation}
where $\epp=\epp(P_2)$ and again we used the transverse polarizations
in the calculation of the form factor and decay width.

\acknowledgements
MD would like to acknowledge the support from the SURA/Jlab fellowship.
This work was supported in part by a grant from the U.S. Department of
Energy (DE-FG02-96ER 40947) and the National Science Foundation
(INT-9906384) and Kyungpook National University Research Fund, 2003.
The North Carolina Supercomputing Center and the National Energy Research 
Scientific Computer Center are also acknowledged for the computing time.

\end{document}